\documentclass[11pt]{article}
 
\usepackage{epsfig}
\begin{document}

\newcommand{\be}{\begin{equation}}
\newcommand{\ee}{\end{equation}}
\newcommand{\bea}{\begin{eqnarray}}
\newcommand{\eea}{\end{eqnarray}}
\newcommand{\da}{\dagger}
\newcommand{\dg}[1]{\mbox{${#1}^{\dagger}$}}
\newcommand{\hlf}{\mbox{$1\over2$}}
\newcommand{\lfrac}[2]{\mbox{${#1}\over{#2}$}}
\newcommand{\scsz}[1]{\mbox{\scriptsize ${#1}$}}
\newcommand{\tsz}[1]{\mbox{\tiny ${#1}$}}

\begin{flushright} 
\end{flushright} 

\begin{center}

\Large{\bf Improving LLR Tests of Gravitational Theory}

\vspace{0.4in}

\normalsize
\bigskip 

James G. Williams$^a$, Slava G. Turyshev$^a$, Thomas W. Murphy, Jr.$^b$\\ 

\normalsize
\vskip 15pt

$^a${\it{Jet Propulsion Laboratory, California Institute of  Technology,\\
Pasadena, CA 91109, U.S.A.}} \footnote{Email: {\tt james.g.williams@jpl.nasa.gov, turyshev@jpl.nasa.gov,  tmurphy@physics.ucsd.edu}} 

\vskip 5pt
$^b${\it{Physics Department, University of California, San Diego, \\CASS-0424, 9500 Gilman Dr., La Jolla, CA 92093, U.S.A.}}


\vspace{0.4in}

\end{center}


\begin{abstract}
 Accurate analysis of precision ranges to the Moon has provided several tests of gravitational theory including the Equivalence Principle, geodetic precession, parameterized post-Newtonian (PPN) parameters $\gamma$  and $\beta$, and the constancy of the gravitational constant {\it G}.  Since the beginning of the experiment in 1969, the uncertainties of these tests have decreased considerably as data accuracies have improved and data time span has lengthened.  We are exploring the modeling improvements necessary to proceed from cm to mm range accuracies enabled by the new Apache Point Observatory Lunar Laser-ranging Operation (APOLLO) currently under development in New Mexico. This facility will be able to make a significant contribution to the solar system tests of fundamental and gravitational physics. In particular, the Weak and Strong Equivalence Principle tests would have a sensitivity approaching 10$^{-14}$, yielding sensitivity for the SEP violation parameter $\eta$ of $\sim 3\times 10^{-5}$, $v^2/c^2$ general relativistic effects would be tested to better than 0.1\%, and measurements of the relative change in the gravitational constant, $\dot{G}/G$, would be $\sim0.1$\% the inverse age of the universe. Having this expected accuracy in mind, we discusses the current techniques, methods and existing physical models used to process the LLR data. We also identify the challenges for modeling and data analysis that the LLR community faces today in order to take full advantage of the new APOLLO ranging station. 

\end{abstract}

\vspace{0.2in}



\section{Introduction}
During more than 30 years of its existence Lunar Laser Ranging (LLR) has become a critical technique available for precision tests of gravitational theory. The 20th century progress in three seemingly unrelated areas of human exploration - quantum physics and optics, astronomy, and human space exploration, led to the construction of this unique interplanetary instrument to conduct very precise tests of fundamental physics in our solar system. LLR offers very accurate laser ranging (currently $\sim$ 2 cm or $\sim 4\times 10^{-11}$ in fractional accuracy) to retroreflectors on the Moon. Analysis of these very precise data contributes to many areas of fundamental and gravitational physics. Thus, these high-precision studies of the Earth-Moon-Sun system provide the most sensitive tests of several key properties of the weak-field gravity, including Einstein's Strong Equivalence Principle (SEP) (the way gravity begets gravity) on which general relativity rests (in fact, LLR is the only means currently available for testing SEP). LLR also provides the strongest limits to date on variability of the gravitational constant (the way gravity is affected by the expansion of the universe), the best measurement of the de Sitter precession rate, and is relied upon to generate accurate astronomical ephemerides. 

In the next few months LLR is poised to take a dramatic step forward, enabled both by detector technology and access to a large-aperture astronomical telescope. The Apache Point Observatory Lunar Laser-ranging Operation (APOLLO) is a unique instrument developed specifically to improve accuracies of LLR ranges to retroreflectors on the Moon. Using a 3.5 m telescope the APOLLO project will push LLR into a new regime of multiple photon returns with each pulse, enabling millimeter range precision to be achieved \cite{[38]}. The project will exploit a large, high-quality modern astronomical telescope at an excellent site to determine the shape of the lunar orbit with a precision of one millimeter. Converting APOLLO's raw range measurements into scientific results requires corresponding improvement in the models used for the data analysis. These improvements, made in parallel with the startup and checkout of the accurate ranging hardware for the Apache Point Observatory, would permit improved solutions for parameters describing the Equivalence Principle, relativity theories, and other aspects of gravitation and solar system dynamics. 

This paper focuses on the current techniques, methods and existing physical models that are used to conduct the LLR tests of $\dot G$ and other PPN parameters.  It is also  discusses the modeling and data analysis improvements necessary to make the order-of-magnitude improvement in these tests that the next-generation of LLR technique enables.  Because of the dramatic increase in the range accuracy anticipated from the APOLLO instrument, today there is an urgent need to develop substantially improved analysis models. Therefore, we consider the challenges for modeling and data analysis needed to take full advantage of this new instrument. The upgraded analysis, combined with APOLLO data, should yield an order-of-magnitude improvement in tests of SEP and the time-variation of Newton's constant, as well as more precise values for the parameterized post-Newtonian (PPN) parameters  $\gamma$ and $\beta$, and for the geodetic precession rate of the lunar orbit.  
This paper addresses these and related issues that are pivotal in taking the full advantage of the newly acquired LLR capabilities.

\section{Lunar Laser Ranging: history and techniques}
  
\subsection{Scientific Background}
\begin{figure}[!t]
    \begin{center} 
\begin{minipage}[t]{.46\linewidth}
  \epsfig{figure=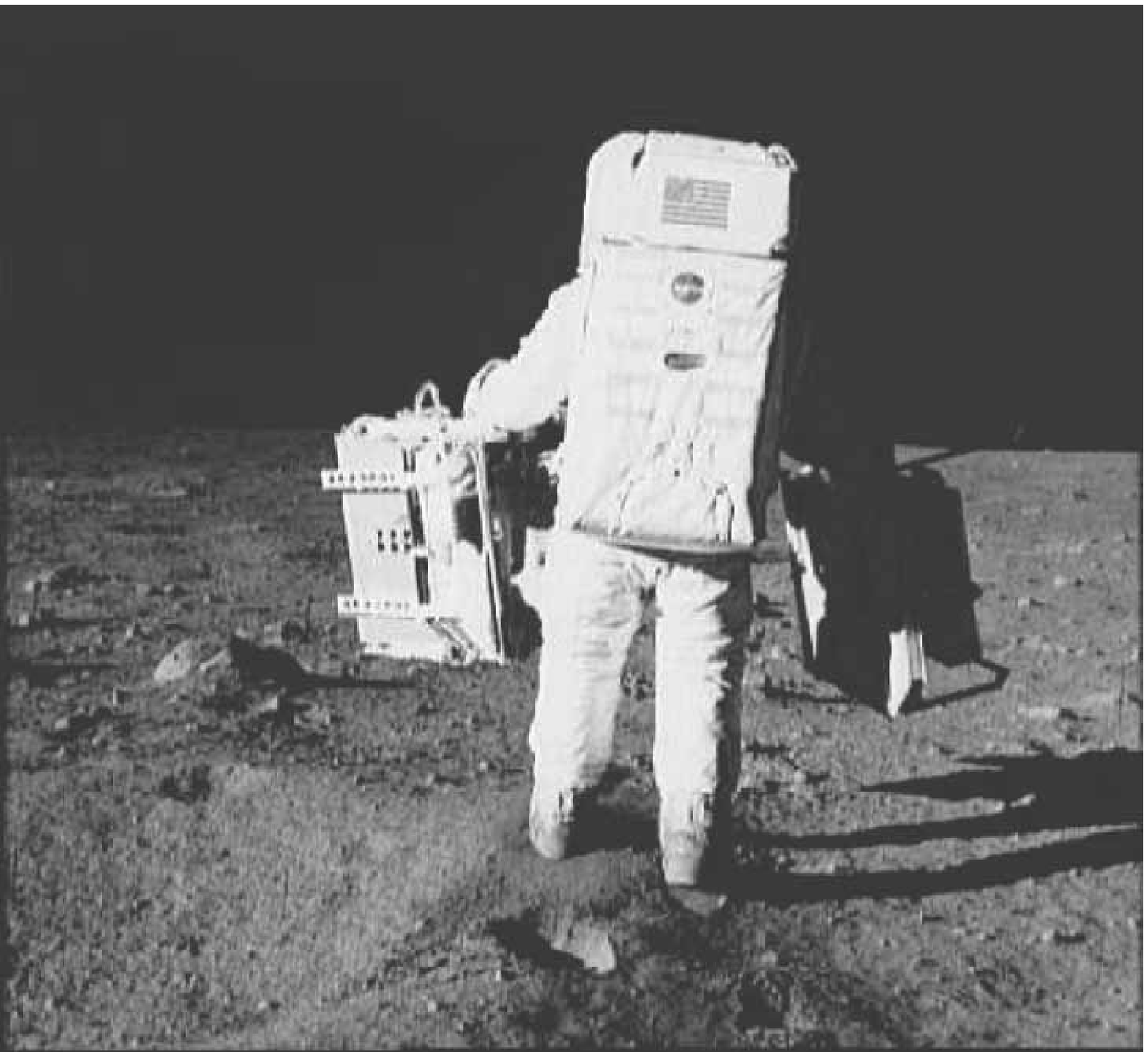,width=78mm} 
    \end{minipage}
\hskip 20pt
\begin{minipage}[t]{.46\linewidth} 
  \epsfig{file=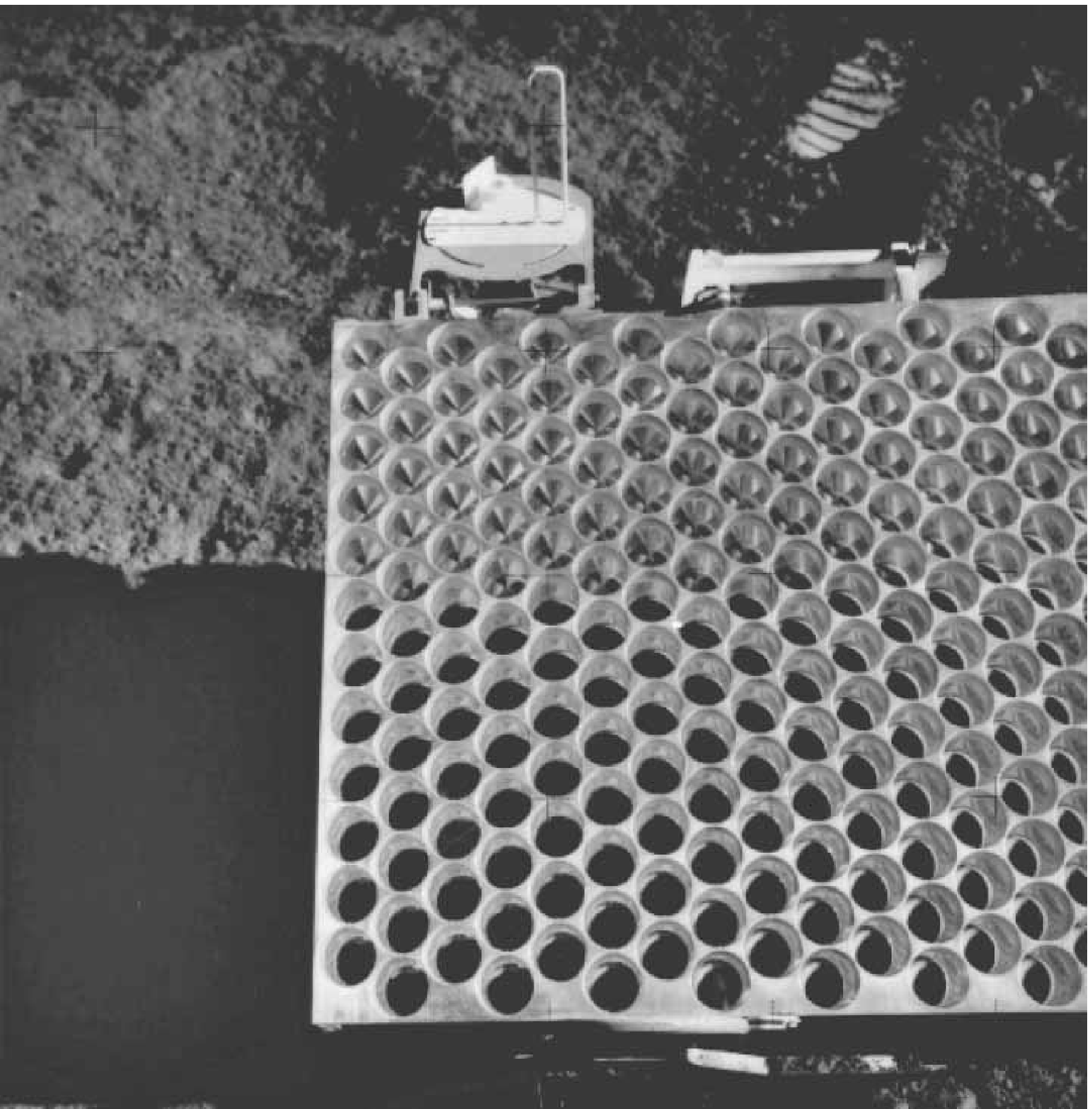,width=71mm}
    \end{minipage}
\caption{(a) The LLR retroreflector being carried across the lunar surface by Buzz Aldrin, an astronaut of the Apollo 11 mission.
(b) The Apollo 15 array of laser retroreflectors.}
 \label{reflectors1}
    \end{center}
\end{figure}

LLR has a distinguished history \cite{[15],[16]} dating back to the placement of retroreflector arrays on the lunar surface by the Apollo 11 astronauts (Figure \ref{reflectors1}a). Additional reflectors were left by the Apollo 14 and Apollo 15 astronauts (Figure \ref{reflectors1}b), and two French-built reflector arrays were placed on the Moon by the Soviet Luna 17 and Luna 21 missions. Figure 2 shows the weighted RMS residual for each year.  Early accuracies using the McDonald Observatory's 2.7 m telescope hovered around 25 cm.  Equipment improvements decreased the ranging uncertainty to $\sim$15 cm later in the 1970s.  In 1985 the 2.7 m ranging system was replaced with the McDonald Laser Ranging System (MLRS).  In the 1980s ranges were also received from Haleakala Observatory on the island of Maui in the Hawaiian chain and the Observatoire de la Cote d'Azur (OCA) in France.  Haleakala ceased operations in 1990.  A sequence of technical improvements decreased the range uncertainty to the current $\sim$ 2 cm.  The 2.7 m telescope had a greater light gathering capability than the newer smaller aperture systems, but the newer systems fired more frequently and had a much improved range accuracy.  The new systems do not distinguish returning photons against the bright background near full Moon, which the 2.7 m telescope could do.  There are some modern eclipse observations.  

LLR accurately measures the time of flight for a laser pulse fired from an observatory on the Earth, bounced off of a corner cube retroreflector on the Moon, and returned to the observatory.  For a general review of LLR see Dickey et al. \cite{[15]}.  A comprehensive paper on tests of gravitational physics is Williams et al. \cite{[19]}. A recent test of the Equivalence Principle is in Anderson and Williams \cite{[21]} and other gravitational physics tests are in Williams et al. \cite{[22]}.  An overview of the LLR gravitational physics tests is given by Nordtvedt \cite{[23]}.  Reviews of various tests of relativity, including the contribution by LLR, are given in Will \cite{[24],[25]}. 

\begin{figure}[!ht]
 \begin{center}
\noindent    
\psfig{figure=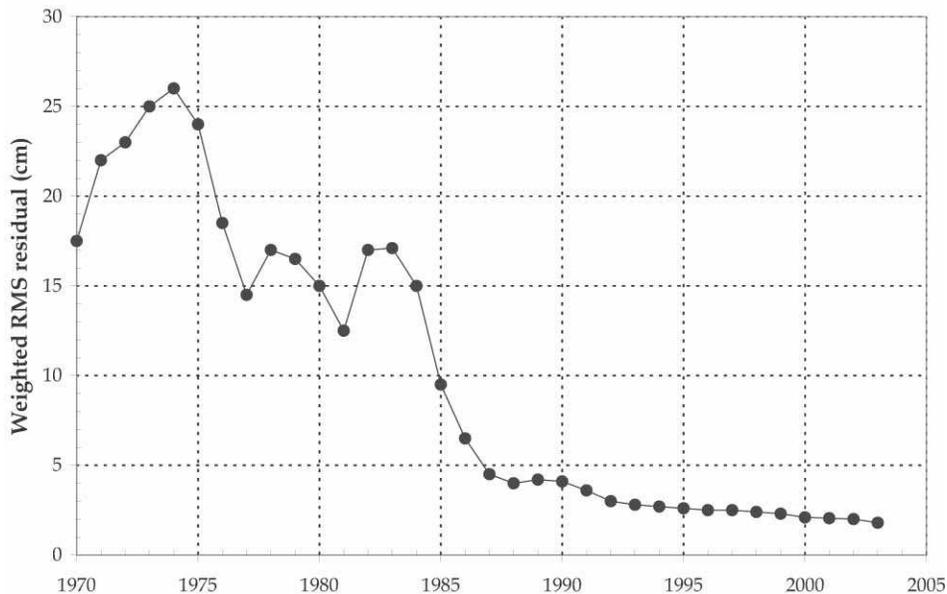,width=125mm}
\end{center}
\vskip -10pt 
  \caption{Historical accuracy of LLR data from 1970 to 2003.  
 \label{accuracy}}
\end{figure} 


The LLR measurements of the past have contributed to a wide range of scientific investigations \cite{[19],[21],[26]}, and are today solely responsible for production of the lunar ephemeris. On the fundamental scientific front, LLR provides the only means for testing SEP--the statement that \emph{all} forms of mass and energy contribute equivalent quantities of inertial and gravitational mass. The SEP is more restrictive than the weak EP, which applies to non-gravitational mass-energy, effectively probing the compositional dependence of gravitational acceleration. It is the equivalence principle which leads to identical accelerations of compositionally different objects in the same gravitational field, and also allows gravity to be viewed as a geometrical property of spacetime--leading to the general relativistic interpretation of gravitation. In addition to the SEP, LLR is capable of measuring the time variation of Newton's gravitational constant, {\it G}, providing the strongest limit available for the variability of this `constant'. LLR can also precisely measure the de Sitter precession--effectively a spin-orbit coupling affecting the lunar orbit in the frame co-moving with the Earth-Moon system's motion around the Sun. The LLR results are also consistent with existence of  gravitomagnetism within 0.1\% of the predicted level \cite{[23],[12n]}, thus making the lunar orbit a unique laboratory for gravitational physics where each term in the relativistic equations of motion was verified to a very high accuracy. Finally, the interior, tidal response, and physical librations (rocking) of the Moon are all probed by LLR, making it a valuable tool for physical selenography \cite{[54]}.

The APOLLO lunar laser-ranging project will yield a one order-of-magnitude improvement in the precision of three important tests of the basic properties of the gravitational interaction.  Below we shall discuss some expected results and their significance for fundamental and gravitational physics.

\subsection{Equivalence Principle Tests}

To date, the experimental evidence for gravitational physics is in agreement with Einstein's Theory of General Relativity (GR).  However, there are a number of theoretical reasons to question the validity of general relativity. Despite the success of modern gauge field theories in describing the electromagnetic, weak, and strong interactions, it is still not understood how gravity should be described at the quantum level. In theories that attempt to include gravity, new long-range forces can arise in addition to the Newtonian inverse-square law. Even at the purely classical level, and assuming the validity of the Equivalence Principle (EP), Einstein's theory does not provide the most general way to generate the space-time metric. Regardless of whether the cosmological constant should be included, there are also important reasons to consider additional fields, especially scalar fields. Although the  scalar fields naturally appear in the theory, their inclusion predicts a non-Einsteinian behavior of the gravitating systems at high energies. These deviations from GR lead to a violation of EP, modification of large-scale gravitational phenomena, and cast doubt upon the constancy of the ``constants''. As a result, this progress has provided new strong motivation for high precision relativistic gravity tests.

The Equivalence Principle, the exact correspondence of gravitational and inertial masses, is a central assumption of general relativity and a unique feature of gravitation. EP tests can therefore be viewed in two contexts: tests of the foundations of the Standard Model of Gravity (i.e. general relativity), or as searches for new physics because, as emphasized by Damour [6], almost all extensions to the Standard Model of particle physics generically predict new forces that would show up as apparent violations of the EP. Easily the most precise tests of the EP are made by simply comparing the free fall accelerations, $a_1$ and $a_2$, of different test bodies, with
\begin{equation}
\frac{\Delta a}{a}\equiv \frac{2(a_1-a_2)}{(a_1+a_2)}=\left(\frac{M_G}{M_I}\right)_1-\left(\frac{M_G}{M_I}\right)_2
\end{equation}
where $M_G$  and $M_I$  represent gravitational and inertial masses of each body. The sensitivity of the EP test is determined by the precision of the differential acceleration measurement divided by the degree to which the test bodies differ (e.g. composition).

\subsubsection{The Weak Equivalence Principle}
The weak form the EP (the WEP) states that the gravitational properties of strong and electro-weak interactions obey the EP. In this case the relevant test-body differences are their fractional nuclear-binding differences, their neutron-to-proton ratios, their atomic charges, etc. General relativity, as well as other metric theories of gravity, predict that the WEP is exact. However, extensions of the Standard Model of Particle Physics that contain new macroscopic-range quantum fields predict quantum exchange forces that will generically violate the WEP because they couple to generalized `charges' rather than to mass/energy as does gravity \cite{[7],[8]}. WEP tests can be conducted with laboratory or astronomical bodies, because the relevant differences are in the test-body compositions.

\subsubsection{The Strong Equivalence Principle}
The strong form of the EP (the SEP) extends the principle to cover the gravitational properties of gravitational energy itself. In other words it is an assumption about the way that gravity begets gravity, i.e. about the non-linear property of gravitation. Although general relativity assumes that the SEP is exact, alternate metric theories of gravity such as those involving scalar fields, and other extensions of gravity theory, typically violate the SEP \cite{[9],[10],[11],[12]}. For the SEP case, the relevant test body differences are the fractional contributions to their masses by gravitational self-energy. Because of the extreme weakness of gravity, SEP test bodies that differ significantly must have astronomical sizes. Currently the Earth-Moon-Sun system provides the best arena for testing the SEP.

To facilitate investigation of a possible violation of the SEP, the ratio between gravitational and inertial masses, $M_G/M_I$  is expressed in the form
\begin{equation}
\frac{M_G}{M_I}=1+\eta\frac{U}{Mc^2}
\end{equation}	  	

\noindent where $U$ is the gravitational self-energy of the body $(U < 0)$, $Mc^2$ is its total mass-energy, and $\eta$  is a dimensionless constant. $U/Mc^2$  is proportional to $M$, so testing the SEP requires bodies the size of the Moon and planets. For the Earth-Moon system,
\begin{equation}
\frac{U_e}{M_ec^2}-\frac{U_m}{M_mc^2}=-4.45\times 10^{-10}
\end{equation}	
where the subscripts $e$ and $m$ denote the Earth and Moon, respectively. Therefore, a violation of the SEP would produce an Earth-Moon differential acceleration of $\Delta a/a =  -4.45\times10^{-10}\eta $.

In general, $\eta$  is a linear function of seven of the ten Parameterized Post-Newtonian (PPN) parameters, but considering only $\beta$  and  $\gamma$
\begin{equation}
\eta=4\beta-\gamma-3
\label{3}
\end{equation}
In general relativity $\eta  = 0$. A unit value for $\eta$  would produce a displacement of the lunar orbit about the Earth \cite{[13],[14]}, causing a 13 meter monthly range modulation.

\subsection{LLR Tests of the Equivalence Principle}
\label{sec:EP}

In essence, LLR tests of the EP compare the free-fall accelerations of the Earth and Moon toward the Sun. Lunar laser-ranging measures the time-of-flight of a laser pulse fired from an observatory on the Earth, bounced off of a retroreflector on the Moon, and returned to the observatory \cite{[15],[16]}. If the Equivalence Principle is violated, the lunar orbit will be displaced along the Earth-Sun line, producing a range signature having a 29.53 day synodic period (different from the lunar orbit period of 27 days). Since the first LLR tests of the EP were published in 1976 \cite{[17],[18]}, the precision of the test has increased by two orders-of-magnitude \cite{[19],[20],[21],[22]}. (Reviews of contributions to gravitational physics by LLR are given by Nordtvedt \cite{[23]} and Will \cite{[24],[25]}.)

From the viewpoint of the EP, the Earth and Moon `test bodies' differ in two significant ways: in composition (the Earth has a massive Fe/Ni core while the Moon has a much smaller core) and in their gravitational self-energies (the Earth is much more massive than the Moon). Therefore, LLR tests the total Equivalence Principle - composition plus self-energy - for the Earth and Moon in the gravitational field of the Sun. Two recent results yield  $\Delta a/a$ values of $(-1 \pm 2)\times 10^{-13}$  \cite{[22]} and $(-0.7 \pm 1.5)\times 10^{-13}$ \cite{[21]}. The latter corresponds to a $2\pm4$ mm amplitude in range.

The LLR result is a null test so it can be argued that it is unlikely that there would be two compensating violations of the Equivalence Principle - composition and self-energy - that essentially cancel. However, because of the fundamental importance of a good SEP test, laboratory tests of the WEP are used to separate with certainty any composition-dependent and self-energy effects. Recent WEP tests performed at the University of Washington (UW) using laboratory test bodies whose compositions are close to those of the actual Earth and Moon set upper limits on any composition-dependent Earth-Moon differential acceleration \cite{[26],[27]}. The random and systematic  $\Delta a/a$ uncertainties of \cite{[27]} are $1.4\pm10^{- 13}$  and  $0.2\times10^{-13}$, respectively. Anderson and Williams \cite{[21]} used the earlier of these WEP results \cite{[26]} to limit the SEP parameter  $\eta = 0.0002\pm0.0008$. If one adopts the more recent WEP test by the UW E\"ot-Wash group \cite{[27]}, one gets an $\eta$ uncertainty of 0.0005. Note that the current intrinsic LLR accuracy, if the WEP were known perfectly, is 0.0003. Therefore, with its 1 mm range accuracy, APOLLO has the capability of determining  $\eta$ to a precision of approximately $3\times 10^{-5}$.

\subsection{LLR Tests of Other Gravitational Physics Parameters}
In addition to the SEP constraint based on Eq.(\ref{3}), the PPN parameters  $\gamma$ and $\beta$  affect the orbits of relativistic point masses, and $\gamma$  also influences time delay \cite{[19]}. LLR tests this $\beta$  and  $\gamma$ dependence, as well as geodetic precession, and $\dot G/G$. Recent work by Williams et al. \cite{[22]} gives uncertainties of 0.004 for  $\beta$ and $\gamma$  deduced from sensitivity apart from   and the SEP, 0.35\% for the geodetic precession, and $1.1\times 10^{- 12}$  yr$^{-1}$ for $\dot G/G$ test.
 
Orbital precession depends on $\beta$  and $\gamma$, so their sensitivity depends on time span of the data. The uncertainty for $\dot G/G$ is improving rapidly because its sensitivity depends on the square of the time span. So 1 mm quality data would improve the $G$ rate uncertainty by an order-of-magnitude in $\sim$ 5 yr while $\gamma$  and geodetic precession would depend on orbital precession time scales: 6.0 yr for argument of perigee, 8.85 yr for longitude of perigee, and 18.6 yr for node.

LLR also has the potential to determine the solar $J_2$ \cite{[22]}, PPN  $\alpha_1$ \cite{[33],[34]}, hunt for influences of dark matter \cite{[13],[35]}, and to test the inverse square law at the scale of $ae \sim$  20,000 km.  A long-range Yukawa interaction has been tested by M\"uller et al. \cite{MG7}.  

\section{Overview of APOLLO}
The Apache Point Observatory Lunar Laser-ranging Operation is a new LLR effort designed to achieve millimeter range precision and corresponding order-of-magnitude gains in measurements of fundamental physics parameters. The APOLLO project design and leadership responsibilities are shared between the University of California at San Diego and the University of Washington. In addition to the modeling aspects related to this new LLR facility, a brief description of APOLLO and associated expectations is provided here for reference. A more complete description can be found in \cite{[38]}.

The principal technologies implemented by APOLLO include a robust Nd:YAG laser with 100 ps pulse width, a GPS-slaved 50 MHz frequency standard and clock, a 25 ps-resolution time interval counter, and an integrated avalanche photo-diode (APD) array. The APD array, developed at Lincoln Labs, is a new technology that will allow multiple simultaneous photons to be individually time-tagged, and provide two-dimensional spatial information for real-time acquisition and tracking capabilities.

The overwhelming advantage APOLLO has over current LLR operations is a 3.5 m astronomical quality telescope at a good site. The site in the Sacramento Mountains of southern New Mexico offers high altitude (2780 m) and very good atmospheric ``seeing'' and image quality, with a median image resolution of 1.1 arcseconds. Both the image sharpness and large aperture enable the APOLLO instrument to deliver more photons onto the lunar retroreflector and receive more of the photons returning from the reflectors, respectively. Compared to current operations that receive, on average, fewer than 0.01 photons per pulse, APOLLO should be well into the multi-photon regime, with perhaps 5-10 return photons per pulse. With this signal rate, APOLLO will be efficient at finding and tracking the lunar return, yielding hundreds of times more photons in an observation than current operations deliver. In addition to the significant reduction in statistical error ($\sim\sqrt{N}$ reduction), the high signal rate will allow assessment and elimination of systematic errors in a way not currently possible.

The principal range signature for the EP tests has a 29.53 day synodic period (lunar phase period the associated argument is called $D$). The current LLR data is not uniformly sampled,   there are no observations near new Moon, and, for recent years, the only observations near full Moon occur during eclipses. The older, less accurate 2.7 m McDonald ranging system was strong enough to get full Moon ranges. APOLLO will also be able to range during full Moon.

There is an important term with half of the synodic period that depends on the product of $G$ and mass, $M$, of the Earth-Moon system. The non-uniform sampling causes a correlation (0.34) between the EP test and $GM$. Nordtvedt has studied the $D$ selection and consequences \cite{[52]}. Reference \cite{[21]} lists several influences on the separation of the EP parameter. APOLLO's data will be taken at lunar phases that will improve the separation of the EP and $GM$ parameters. In addition, APOLLO plans to incorporate a precision superconducting gravimeter and a GPS receiver that will measure crustal motions at the telescope site with millimeter precision. This information will significantly enhance our ability to model site displacements from tides, atmosphere and ocean loading, and ground-water infusion.

\subsection{APOLLO Contribution to the Tests of Gravity}

The new LLR capabilities offered by the newly developed APOLLO instrument offer a unique opportunity to improve accuracy of a number of fundamental physics tests. Some of them would have a profound effect on our understanding of the evolution of our universe. Thus, in the search for experimentally testable predictions of the scalar-tensor theories of gravity, Damour and Nordtvedt \cite{[28]} recently considered the evolution of tensor-scalar theories following the inflation era. They conclude that the coupling between the scalar and tensor fields weakens with time so that GR is approached, but not reached. They consider a range of possibilities resulting in a deviation from GR described by $\sim 10^{-7} < \eta  < 10^{- 4}$. Additionally, Damour and Nordtvedt note that there exists a connection between   $\dot G/G$ and $\eta$. If {\it G} changes at a rate comparable to the reported change in the fine structure constant ($\dot\alpha/\alpha \sim 10^{-15}$ yr$^{- 1}$) \cite{[29]}, $\eta$  would be approximately $10^{- 5}$. Thus, an order-of-magnitude LLR range improvement would give an   uncertainty within reach of the predictions by Damour and Nordtvedt, and comparable to the value implied by $\dot\alpha.$

It is already known via LLR that the Moon's orbit is {\it not} displaced by a violation of the SEP at the $\sim$ 4 mm level of precision--based on 2 cm raw range precisions--constraining the SEP violation parameter $\eta$ at the level of about 5 parts in 10$^4$ (see Section \ref{sec:EP}). The high quality data from APOLLO should permit sub-millimeter determinations of the SEP-induced polarization of the lunar orbit, thereby achieving a few parts in 10$^5$ precision on the potential SEP violation. Various alternative theories of physics
(string, or M-theory, for example) predict new particles such as the dilaton and moduli that couple with gravitational strength and whose exchange violates the SEP.

The SEP relates to the non-linearity of gravity (how gravity affects itself), with the PPN parameter  $\beta$ representing the degree of non-linearity. Thus LLR provides the best way to measure  $\beta$, as suggested by the strong dependence of $\eta$ on  $\beta$  in Eq. (\ref{3}). The parameter $\gamma$  has been measured independently via time-delay and gravitational ray-bending techniques. The published Viking \cite{[29]} and Very Long-Baseline Interferometry (VLBI) \cite{[31],[32]} uncertainties for $\gamma$  are 0.002, 0.002, and 0.0022, respectively. Combining the above limits on  $\eta$  from LLR and laboratory WEP tests with the Viking and VLBI results for  $\gamma$  gives  $|\beta-1|<0.0005$, the limit given by \cite{[21]}. The uncertainty in $\beta$  determined in this way is dominated by the uncertainty in $\gamma$. A much more accurate result for $\gamma$ is expected from the Cassini conjunction experiment \cite{cassini} and would lead to a significant improvement of parameter $\beta$ determination.

The APOLLO project will push LLR into the regime of millimetric range precision which translates to an order-of-magnitude improvement in the determination of fundamental physics parameters. For the Earth and Moon orbiting the Sun, the scale of relativistic effects is set by the ratio $(GM / r c^2)\sim v^2 /c^2 \sim 10^{-8}$.  Relativistic effects are small compared to Newtonian effects.  The Apache Point 1 mm range accuracy corresponds to $3\times 10^{-12}$ of the Earth-Moon distance.  The resulting LLR tests of gravitational physics would improve by an order of magnitude: the Equivalence Principle would give uncertainties approaching $10^{-14}$, tests of general relativity effects would be $<0.1$\%, and estimates of the relative change in the gravitational constant would be 0.1\% of the inverse age of the universe. This last number is impressive considering that the expansion rate of the universe is approximately one part in 10$^{10}$ per year.

Therefore, the gain in our ability to conduct even more precise tests of fundamental physics is enormous, thus this new instrument stimulates development of better and more accurate models for the LLR data analysis at a mm-level. In the next section we shall discuss the current state and the needed future improvements in these models in more details. 

\section{Current Formulation of Range Model}

LLR measures the range from an observatory on the Earth to a retroreflector on the Moon. The center-to-center distance of the Moon from the Earth, with mean value 385,000 km, is variable due to such things as eccentricity, the attraction of the Sun, planets, and the Earth's bulge, and relativistic corrections. In addition to the lunar orbit, the range from an observatory on the Earth to a retroreflector on the Moon depends on the positions in space of the ranging observatory and the targeted lunar retroreflector. Thus, orientation of the rotation axes and the rotation angles of both bodies are important with tidal distortions, plate motion, and relativistic transformations also coming into play. To extract the gravitational physics information of interest it is necessary to accurately model a variety of effects. 

The existing model formulation, and its computational realization in computer code, is the product of many years of effort. Most of the sensitivity to gravitational physics parameters is through the orbital dynamics, though PPN  $\gamma$ also influences relativistic time delay. The successful analysis of LLR data requires attention to geophysical and rotational effects for the Earth and the Moon in addition to the orbital effects. The existing model and computer codes fit the recent range data with a root-mean-square (RMS) residual of 17 mm (150 mm in the one-way distance corresponds to 1 ns in two-way time of flight). A number of  few-millimeter-sized effects are not presently modeled, and we plan to carry out such improvements in model accuracy. This effort should improve both random and systematic influences on the fits. 

\subsection{Existing Range Model}

In the current model the time-of-flight (`range') calculation consists of the round-trip `light time' from a ranging site on the Earth to a retroreflector on the Moon and back to the ranging site. The vector equation for the one-way range vector  $\vec\rho$  is:
 \begin{equation}
\vec\rho=\vec r-\vec R_e+\vec R_m
\end{equation}
\noindent where $\vec r$  is the vector from the center of the Earth to the center of the Moon, $\vec R_e$  is the vector from the center of the Earth to the ranging site, and $\vec R_m$ is the vector from the center of the Moon to the retroreflector array. The times at the Earth and Moon are different: transmit, bounce, and receive times. Due to the motion of the bodies the light-time computation is iterated for both the up and down legs. The orbits and lunar rotation come from the numerical integration. The solar system barycenter is the coordinate frame for relativistic computations. The total time of flight is the sum of the up and down leg geometry plus delays due to atmosphere and relativity.

It is convenient to represent the model as a sum of its major components, including the description of the dynamics of the lunar orbit and rotation, the Earth's and Moon's finite size effects as well as the relativistic corrections. For illustrative purposes we shall discuss each of the components separately, in particular the existing range model accounts for:

\subsubsection{Orbit Dynamics \& Lunar Rotation}
The lunar and planetary orbits and the lunar rotation result from numerical integration of the differential equations of motion. 
The existing model for orbital motion accounts for:
\begin{itemize}
\item
Newtonian and relativistic point mass gravitational interaction between Sun, Moon, and nine planets. This is parameterized for the tests of EP, $\dot G$ and PPN  $\beta$  and  $\gamma$.

\item Newtonian attraction of the largest asteroids.

\item Newtonian attraction between a point mass and a body with gravitational harmonics: Earth ($J_2, J_3, J_4$), Moon (2-nd through 4-th degree spherical harmonics), and Sun ($J_2$).

\item Attraction from tides on Earth and Moon including both elastic and dissipative components.
\end{itemize}

Numerical integration of the rotation of the Moon accounts for:
\begin{itemize}
\item	Torques from point mass attraction of Earth, Sun, Venus and Jupiter. The lunar gravity field is included with its spherical harmonics of 2-nd through 4-th degree.
\item	Figure-figure torques between Earth ($J_2$) and Moon ($J_2$ and $C_{22}$).
\item	Torques from tides raised on the Moon include elastic and dissipative components (a time delay formulation) \cite{[41],[42]}.
\item	The fluid core of the Moon is considered to rotate separately from the mantle. A dissipative torque at the lunar solid-mantle/fluid-core interface couples the two  \cite{[42]}. The rotations of both mantle and core are integrated.
\end{itemize}

\subsubsection{Finite-size Earth Effects}
\begin{itemize}
\item The ranging station coordinates include rates for plate motion.
\item Tidal displacements on Earth are scaled by terrestrial Love numbers $h_2$ and $l_2$ with a core-flattening correction for a nearly diurnal term. There is also a ``pole tide'' due to the time-varying part of the spin distortion.
\item The orientation of the Earth's rotation axis includes precession and nutation. The body polar axis is displaced from the rotation axis by polar motion.  The rotation includes UT1 variations. 
\item Atmospheric time delay follows \cite{[39]}. It includes corrections for surface pressure, temperature and humidity which are measured at the ranging site.
\end{itemize}
A useful reference for Earth effects is \cite{[43]}.
\subsubsection{Finite-size Moon Effects}
\begin{itemize}
\item	The Moon-centered coordinates of the retroreflectors  are adjusted for tidal displacements on the Moon. Tides are scaled by the lunar Love numbers $h_2$ and $l_2$.
\item	The rotation matrix for the Moon depends on three Euler angles that come from the numerical integration.
\end{itemize}
 
\subsubsection{Relativistic Corrections}
The frame for computations is the solar system barycenter and relativistic corrections are needed to convert that position and time to and from geocentric position and station time.  There are also light-time delays due to gravity.
\begin{itemize}
\item	The motion of the Earth and Moon with respect to the solar system barycenter requires a Lorentz contraction for geocentric ranging station and Moon-centered reflector coordinates.
\item	In the relativistic time transformation there are time-varying relativistic terms due to the motion of the Earth's center with respect to the solar system barycenter.
\item	The displacement of the ranging station from the center of the Earth causes relativistic time variations.
\item	The propagation of light in the gravity fields of the Sun and Earth causes a relativistic time delay.
\end{itemize}

\subsection{Fit Parameters,  Partial Derivatives, and Computation}

For each fit parameter there must be a partial derivative of the ``range'' with respect to that parameter. The partial derivatives may be separated into two types: geometrical and dynamical. 
Geometrical partials are explicit in the model for the time of flight. Examples are partial derivatives for geocentric ranging station coordinates, Moon-centered reflector coordinates, station rates due to plate motion, tidal displacement Love numbers $h_2$ and $l_2$, nutation coefficients, diurnal and semidiurnal UT1 coefficients, angles and rates for the Earth's orientation in space, ranging biases and many more.

Dynamical partials are for parameters that enter into the model for numerical integration of the orbits and lunar rotation. Examples are dynamical partial derivatives for the masses and initial conditions for the Moon and planets, the masses of several asteroids, the initial conditions for the rotation of both the lunar mantle and fluid core, Earth and Moon tidal parameters, lunar moment of inertia combinations $(B-A)/C$ and $(C-A)/B,$ lunar third-degree gravity field, a lunar core-mantle coupling parameter, Equivalence Principle $M_G/M_I$, PPN parameters $\beta$  and $\gamma$, geodetic precession, solar $J_2$, and a rate of change for the gravitational constant $G$. Dynamical partial derivatives for the lunar and planetary orbits and the lunar rotation are created by numerical integration. At the time of the range calculation a file of integrated partial derivatives for orbits and lunar rotation with respect to dynamical solution parameters is read and converted to partial derivatives for range with respect to those parameters. The partial derivative for PPN $\gamma$  has both dynamical and geometrical components. Useful references may be found in \cite{[19],[41],[42],[40]}.

The analytical model has its computational realization in a sequence of computer programs. Briefly these a) numerically integrate lunar and planetary orbits, lunar rotation, and dynamical partial derivatives, b) compute the model range for each data point, form the pre-fit residual, and compute range partial derivatives, c) solve the least-squares equations, and d) generate and plot post-fit residuals. All these programs would have to be updated in order to accommodate the new accuracy of the LLR data expected with the APOLLO instrument.  

\section{Next Steps and Conclusions}

In order to respond to the technological challenge presented  by the APOLLO project, there is an urgent need to improve the modeling accuracy by more than an order-of-magnitude in order to match the millimeter quality range accuracy expected from APOLLO. This requires investigating a number of effects that are ignored in the current model. Some effects may be precisely calculated while others require adjustable solution parameters.  

Planning for model improvement to accompany the high-quality range data is already underway. We have begun exploring components of the current model which deserve attention. The high-precision ranging data will help pinpoint deficiencies in the model, ultimately enabling the model to reach mm precision or better.

There are a number of small influences on the Earth, Moon, and orbit.
The larger effects generally have higher priority, but there are subtleties. Most serious are coherent effects that can mimic phenomena of interest. Thus, the effect of solar radiation pressure influences the orbit of the Moon.  This causes a 3.7 mm synodic monthly periodicity which mimics the largest EP signature \cite{[51]}. Additionally, the monthly thermal expansion cycle of retroreflector array heights is 1-2 mm for the Apollo retroreflector arrays which sit on the soil, but is at least 5 mm for the Lunokhod reflector which is $\sim$1 m above the surface.  The thermal cycle also occurs at the synodic period which is important to the EP test. Solar tides on the Moon ($\sim 2$ mm) occur at half the synodic period. 

There are also two interesting effects influencing $\dot G/G$ tests (current uncertainty $10^{-12}$ yr$^{-1}$).  For the Sun, changes in $G$ and $M$ cannot be distinguished.  The solar mass loss relative to its mass is $-7\times10^{-14}$ yr$^{-1}$ and can be accurately computed.  For accuracies below $10^{-14}$ yr$^{-1}$ thermal reradiation from the Earth and Moon should be considered.  At present, the integration of the rotation of the Moon is Newtonian.  There is an annual relativistic term of 8 mm amplitude at the equator, due to the time transformation, which would project into 3 mm or less in range (the projection depends on the reflector location).  Also in the lunar rotation, the geodetic ``precession'' will alter the tilt rather than the precession rate because the precessing equator is a forced motion.

For near future APOLLO data analysis, coherent effects need to be modeled with precisions better than 1 mm.\footnote{It may be argued that, in order to achieve the 1 mm aggregate RMS, a modeling accuracy considerably better than the expected range accuracy of the instrument is justified for individual effects.}  The least trouble is caused by random noise and effects that act like random noise since large numbers of observations reduce their influence. An example is atmospheric variations that are unrelated to lunar position. Many effects are in between these extremes. They may have modest correlations with some solution parameters, but low correlation with the parameters of interest. So both size and signature determine the modeling priority.
  
The expected improvement in the accuracy of LLR tests of gravitational physics expected with the new APOLLO instrument will bring significant new insights to our understanding of the fundamental physics laws that govern the evolution of our universe. The scientific results are significant which justify the more than 35 years of history of LLR research and technology development. It also imposes a need to develop the accurate range models and data analysis techniques that will sustain this new technological challenge.  

In conclusion, LLR provides the most precise way to test the EP for gravity itself, the best way to test for non-gravitational long-range fields of dark matter as well as for time variation of Newton's constant. With technology improvements and substantial access to a large-aperture, high-quality telescope, the APOLLO project will take full advantage of the lunar retro-reflectors and will exploit the opportunity provided by the unique Earth-Moon ``laboratory'' for fundamental gravitational physics.


\section*{Acknowledgments}
The authors would like to thank William Folkner of JPL for suggesting the thermal investigation. The work described here was carried out at the Jet Propulsion Laboratory, California Institute of Technology, under a contract with the National Aeronautics and Space Administration.



\end{document}